\documentclass[doublecol,linenumbers]{epl2}
\usepackage{booktabs}

\title{Critical size of ego communication networks}
\shorttitle{Critical size of ego communication networks}

\author{Qing Wang\inst{1,2} \and Jian Gao\inst{3} \and Tao Zhou\inst{3,4}\footnote{E-mail: zhutou@ustc.edu} \and Zheng Hu\inst{1,2}\footnote{E-mail: huzheng@bupt.edu.cn} \and Hui Tian\inst{1,2}}
\shortauthor{Q. Wang \etal}
\institute{
  \inst{1} State Key Laboratory of Networking and Switching Technology, Beijing University of Posts and Telecommunications, Beijing 100876, PRC\\
  \inst{2} School of Information and Communication Engineering, Beijing University of Posts and Telecommunications, Beijing 100876, PRC\\
  \inst{3} CompleX Lab, Web Sciences Center, University of Electronic Science and Technology of China, Chengdu 611731, PRC\\
  \inst{4} Big Data Research Center, University of Electronic Science and Technology of China, Chengdu 611731, PRC\\
}


\abstract{
With the help of information and communication technologies, studies on the overall social networks have been extensively reported recently. However, investigations on the directed Ego Communication Networks (ECNs) remain insufficient, where an ECN stands for a sub network composed of a centralized individual and his/her direct contacts. In this paper, the directed ECNs are built on the Call Detail Records (CDRs), which cover more than 7 million people of a provincial capital city in China for half a year. Results show that there is a critical size for ECN at about 150, above which the average emotional closeness between ego and alters drops, the balanced relationship between ego and network collapses, and the proportion of strong ties decreases. This paper not only demonstrate the significance of ECN size in affecting its properties, but also shows accordance with the ``Dunbar's Number". These results can be viewed as a cross-culture supportive evidence to the well-known Social Brain Hypothesis (SBH).
}
\begin{document}
\maketitle

\section{Introduction}
The widespread of information and communication technologies makes it convenient for people to communicate with each other. In the meantime, the social behaviors of users are much easier to be recorded precisely. The accumulation of such digital communication records provides a new approach to study social networks, human dynamics and other related areas \cite{Borgatti2009,Zhao2011}. For example, the Call Detail Records (CDRs) can be used to study the human communication behavior and human mobility \cite{Fisher2005,Gonzalez2008,Song2010,Miritello2011,Toole2015}. Besides, the digital communication records are perfect for analysing Ego Networks (ENs), which examine the ties connecting the target individual (ego) and his/her direct contacts (alters) \cite{Roberts2011}. The studies on the ENs can benefit us in understanding how human organize their personal social networks \cite{Arnaboldi2012}. Here, we focus on the Ego Communication Network (ECN) from CDRs.

Most literatures focus on the structure and function of the overall mobile communication networks \cite{Wu2010,Onnela2007,Nanavati2008,Eagle2009,Jiang2013}. For example, Onnela \etal \cite{Onnela2007} uncovered the existence of the weak tie effects and further demonstrated its significance to the network's structural integrity by doing comprehensive analyses of the weighted mobile communication networks. Nanavati \etal \cite{Nanavati2008} revealed the Bow-Tie shape of the mobile communication networks and proposed a Treasure-Hunt model to picture the whole graph of telecom users according to the in-degrees and out-degrees. Eagle \etal \cite{Eagle2009} found that it is possible to infer $95\%$ of friendships accurately based only on the mobile communication data in comparison with the self-report survey data. Wu \etal \cite{Wu2010} revealed the bimodal distribution in human communication, which is composed of a generic Poisson process in individual human behavior and a power-law-like bursts through the interactions with other individuals. Furthermore, Jiang \etal \cite{Jiang2013} argued that the inter-call duration follows a power law distribution with an exponential cutoff at the population level and Weibull distribution at the individual level. As illustrated above, much attentions have been paid to uncovering the overall features of mobile communications networks.

Apart from the overall analyses, some researchers also considered the mobile communication networks from ego perspective \cite{Miritello2013,Miritello2013b,Saramaki2014}. Miritello \etal \cite{Miritello2013} uncovered the time constraints and communication capacity by studying the individual's communication strategies. Saram{\"a}ki \etal \cite{Saramaki2014} showed that individuals have robust and distinctive social signatures persisting overtime.

Ego network has long been a interesting topic in social science. For example, Dunbar \cite{Dunbar1998} proposed the Social Brain Hypothesis (SBH), which implies that the constraint on the group size of primates arises from the information-processing capacity, being highly correlated with the neocortex size \cite{Dunbar1992}. As to human society, the SBH suggests that the people can only maintain about 150 friends due to the cognitive constraints \cite{Dunbar2010,Ruiter2011}. The number 150 is thus well known as ``Dunbar's Number".

The results from the extensive studies of online social networks turned out supporting the ``Dunbar's Number" \cite{Arnaboldi2013,Dunbar2015,Zhao2014,Gonccalves2011,Guo2013}, and they can also be seen as supportive evidences to the SBH theory. Arnaboldi \etal \cite{Arnaboldi2013} argued that the new technologies could not alter the way people organize their social networks by analyzing the Facebook and Twitter networks. Dunbar \etal \cite{Dunbar2015} concluded that the structure of online social networks mirrors those in the off-line world. Zhao \etal \cite{Zhao2014} found that there is still a magic upper limit of social relationships at $200\sim300$ for Facebook users. Moreover, Gon{\c{c}}alves \etal \cite{Gonccalves2011} argued that Dunbar's Number is $100\sim200$ in Twitter networks. Guo \etal \cite{Guo2013} showed that one could only maintain relationships with $65$ people in a single city based on the study of the New Orleans' Facebook data.

Among all these diverse literatures, most studies paid attention to either the properties of the overall network or some particular features of the undirected ECNs, for example the weight of the undirected link. In fact, ECNs are intrinsically directed, and different communication directions have different social meanings. How to characterize these differences and what are the correlations between these different ECN properties and ECN sizes still need further investigations.

In this letter, we study how the directed ECN properties are affected as the expanding of ECNs. Here ECNs are built on CDR datasets, covering over 7 million people of a provincial capital city in China for half a year. We mainly focus on the average node strength, the attractiveness balance and tie balance of directed ECNs in the analysis. Empirical results show that once the ECN size exceeds a critical value at about 150, the average emotional closeness drops, the attractiveness between ego and the network changes significantly and the proportion of strong ties decreases. Interestingly, the critical sizes obtained in this paper are very close to the well-known ``Dunbar's Number" 150, which also provides a supportive evidence to the SBH theory from a different cultural background.

\section{Methods}
Mobile communication is useful in maintaining social relationships \cite{Saramaki2014,Zhou2005}. Different from the reciprocity nature of communication in life, mobile communication is intrinsically directed. The directions of communication are crucial in understanding the relationships among people and the information diffusion process especially in the cyber social networks \cite{Brzozowski2011}. For example, when A makes a call to a new contact B, it means A intends to spend resources (e.g., time, money and energy) on maintaining the social relationship with B. In return, B can either reply to A to establish a reciprocal relationship or just ignore it. Thus the total number of out-contacts can reflect the attractiveness of a network (represented by A's out-contacts including B) to an ego (A), while the total number of in-contacts can tell the attractiveness of an ego (B) to the network (represented by B's in-contacts including A). As a result, the network directions can be used to infer the directed attractiveness between an ego and the network. No matter strong or weak, social relationships can always help people to survive by providing material support or information \cite{Brown1987,Holt2010,Zhou2005}.

Based on the above considerations, we model the mobile communication network as a directed weighted graph $G(V,E)$ with the number of nodes and links being $|V|=N$ and $|E|=L$, respectively. Here nodes represent mobile users and the directed links represent the social relationships among users. There are no self-connections and multiple links in this network. For a directed link $l_{ij}$ from user $i$ to user $j$, the link weight is denoted as $w_{ij}$, which is the number of calls that user $i$ has made to user $j$. According to the directions of communication, the alters are divided into two sets for each ego, namely, the in-contact set $C_{i}^{in}$ and the out-contact set $C_{i}^{out}$. The sizes of $C_{i}^{in}$ and $C_{i}^{out}$ are in-degree $k_{i}^{in}$ and out-degree $k_{i}^{out}$, respectively. $k_{i}^{out}$ represents the ECN size ego $i$ maintains, and $k_{i}^{in}$ can reflect the influence of ego $i$. In this paper, we mainly focus on $k^{out}$, because it represents the number of alters an ego intends to spend resources to maintain.

Link weight can reflect the emotional closeness between mobile users \cite{Saramaki2014,Zhou2005}. In order to better observe the relationship between ego and alters, the average node weight $\overline{w_{i}}$ is introduced, which is defined as
\begin{equation}\label{formulaaw}
\overline{w_{i}} = \frac{1}{k_{i}^{out}} \sum_{j \in C_{i}^{out}} {w_{ij}},
\end{equation}
where $w_{ij}$ is the weight of link $l_{ij}$, and $k_{i}^{out}$ is the ECN size. $\overline{w_{i}}$ indicates the average emotional closeness between an ego and the alters.

Starting from the definition of $C_{i}^{in}$ and $C_{i}^{out}$, two basic questions arise. One is what the relationships between egos and the network are, i.e., most of the users tend to reach out for others or do they wait for being contacted. In order to capture such relationships, we introduce attractiveness balance $\eta$. $\eta_{i}$ for ego $i$ is defined in a straightforward way:
\begin{equation}\label{formulaeta}
\eta_{i}=\frac{k_{i}^{in}}{k_{i}^{out}}.
\end{equation}
The attractiveness balance $\eta = 1$ indicates that the number of relationships from the network to an ego and the number of relationships from the ego to the network are the same, suggesting the balance of the attractiveness. Larger $\eta$ implies the stronger attractiveness of an ego, whilst smaller $\eta$ refers to a weaker attractiveness.

The directions of network also distinguish the bidirectional links (alters appear in both $C^{in}$ and $C^{out}$) from the unidirectional links (alters only appear in either $C^{in}$ or $C^{out}$). The reciprocal relationships are stronger than the unidirectional relationships, thus they can be viewed as strong ties and weak ties. Without loss of generality, strong ties in this paper suggest reciprocal intentions of forming the relationships thus have larger chance to provide mutual support than the weak ties.

The second basic question is how people manage the proportion of strong and weak ties within their ECNs. In order to answer this question, we introduce the tie balance $\theta$, which is defined as the proportion of the bidirectional alters. Mathematically, it reads:
\begin{equation}\label{formulatheta}
\theta_{i}=\frac{|C_{i}^{in} \cap C_{i}^{out}|}{|C_{i}^{in} \cup C_{i}^{out}|}.
\end{equation}
In fact, $\theta_{i}$ is the Jaccard distance \cite{Levandowsky1971} between $C_{i}^{in}$ and $C_{i}^{out}$. $\theta = 1$ means the ego only have bidirectional alters and $\theta = 0$ means the ego only have unidirectional social relationships. Most people can maintain a comparatively stable $\theta$ so that they can preserve the balance between the strong and weak ties within their ECNs.

To study the impacts of ECN size on its properties, we investigate the changes of correlations between the ECN properties ($\overline{w}$, $\eta$ and $\theta$) and the ECN size $k^{out}$. Critical size of ECN ($k_{c}^{out}$) corresponds to the ECN size where such correlations show different patterns for smaller and larger ECNs. In order to determine the exact value of $k_{c}^{out}$, we calculate the absolute value of Pearson correlation \cite{Pearson1895} between ECN properties and the ECN size. As $k_{i}^{out}$ increases, the correlation $\rho_{i}$ between the ECN sizes that are smaller than $k_{i}^{out}$ and the corresponding ECN properties is calculated. The extremum value of $\rho$ always indicates the transformation of the correlation between the ECN property and ECN size. The social meanings of the critical sizes obtained from $\overline{w}$, $\eta$ and $\theta$ will be detailed in the empirical results.

In preliminary summary, all the introduced metrics are derived from the in-contact set $C^{in}$ and out-contact set $C^{out}$ for each ego $i$ based on the directed nature of communication networks. Among them, $\overline{w}$ characterize the average strength of relationship between ego and alters in $C^{out}$. Both $\eta$ and $\theta$ describe the relationships between $C^{in}$ and $C^{out}$. $\eta$ directly measures the size balance between $C^{in}$ and $C^{out}$ whilst $\theta$ measures the Jaccard distance between these two sets. Further, we propose a method to identify the critical size of ECN by calculating the correlations between the ECN properties and the ECN size.

\begin{table}
\centering
    \caption{Basic statistics of the mobile communication networks. $N_{t}$ is the number of \emph{total} users. $N_{l}$ is the number of the \emph{local} users. $L_{t}$ is the number of \emph{total} links.}
    \begin{tabular*}{0.48\textwidth}{@{\extracolsep{\fill}}c|ccc}
    \toprule
    Time & $N_{t}$ & $N_{l}$ & $L_{t}$  \\
    \midrule
    Jan. & 6520121 & 751643 & 32521180  \\
    Feb. & 6234877 & 742504 & 27600221  \\
    Mar. & 6481767 & 783751 & 32720452  \\
    Apr. & 6526250 & 777486 & 32383231  \\
    May. & 6561107 & 787614 & 34119390  \\
    Jun. & 6531076 & 787156 & 33461297  \\
    \bottomrule
    \end{tabular*}
\label{Tab1}
\end{table}

\begin{figure*}[!t]
\centering
\includegraphics[width=0.9\textwidth]{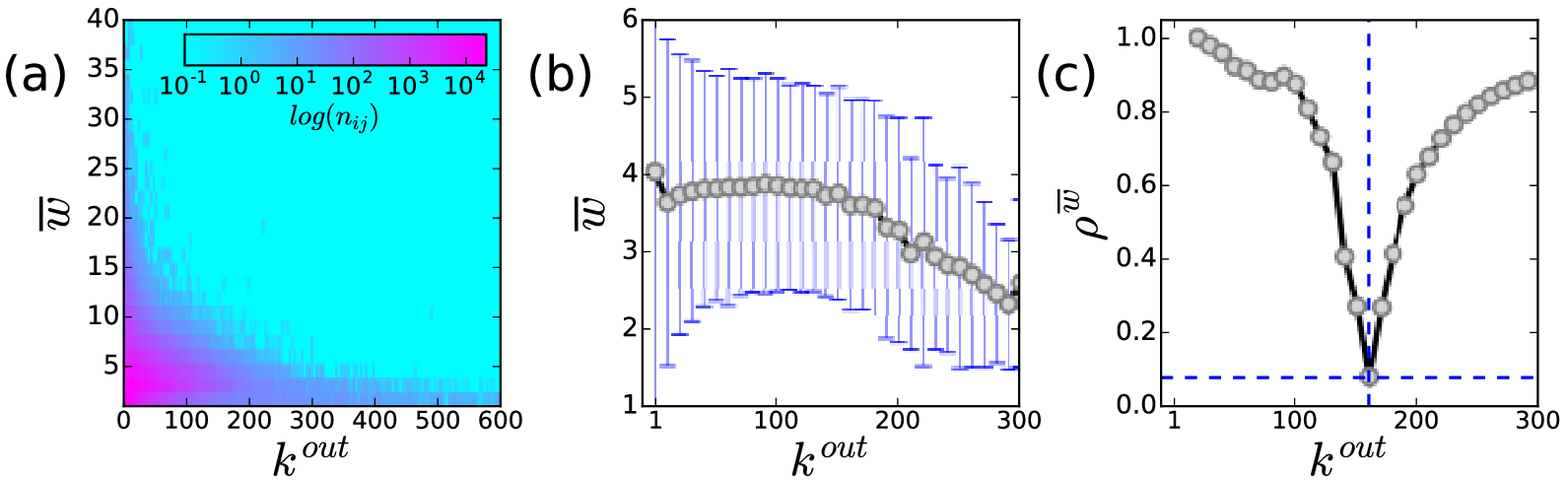}
\caption{(Color online) 2-dimensional distribution and the correlation between average node weight $\overline{w}$ and ECN size $k^{out}$. (a) 2-dimensional distribution for $\overline{w}$ and $k^{out}$. The color represents the number of egos ($n_{ij}$) in log scale. (b) Correlation between $\overline{w}$ and $k^{out}$. The data are divided into bins with length 10 and the average value in of each bin is presented. The error bar corresponds to the standard deviation. (c) The critical size $k_{c}^{out}$ is identified as 161, where $\rho^{\overline{w}}$ reaches its minimum value.}
\label{Fig1}
\end{figure*}

\section{Data description}
The anonymous CDRs are collected by mobile operators for billing and network traffic monitoring. The basic information of the data set contains the anonymous IDs of callers and callees, time stamps, call durations, and so on. In this study, the data set is furnished by one of the largest mobile operators in China. And it covers 7 million people of a provincial capital city in China for half a year from Jan. to Jun., 2014. According to the operator a user chooses, all the users within this data set can be divided into two categories, namely, the $local$ users (customers of the mobile operator who provide this data set) and the $alien$ users (customers from another mobile operators). The reason for such distinction is that only a fraction of $alien$ users' communication records are included within this data set, $i. e.$ only local users have enough data for building their ECNs. As a result, we will derive the metrics of the users based on the whole data set, while only focus on $local$ users in the analyses. The basic statistics of the mobile communication networks are summarized in table \ref{Tab1}.
\section{Empirical results}
\begin{figure*}[!t]
\centering
\includegraphics[width=0.9\textwidth]{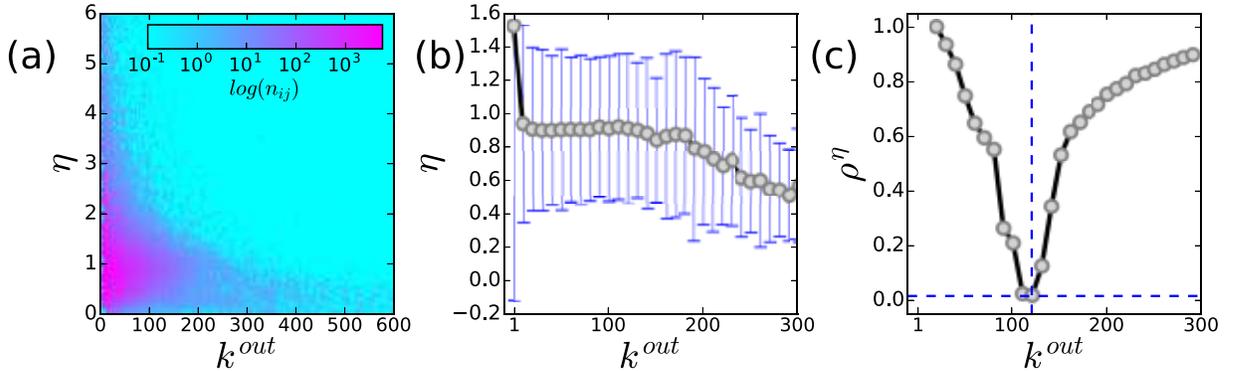}
\caption{(Color online) 2-dimensional distribution and the correlation between attractiveness balance $\eta$ and ECN size $k^{out}$. (a) 2-dimensional distribution for $\eta$ and $k^{out}$. The color represents the number of egos ($n_{ij}$) in log scale. (b) Correlation between $\eta$ and $k^{out}$. The data are divided into bins with length 10 and the average value of each bin is presented. The error bar corresponds to the standard deviation. (c) The critical size $k_{c}^{out}$ is identified as 121, where $\rho^{\eta}$ reaches its minimum value.}
\label{Fig2}
\end{figure*}

In this section, we mainly focus on the average node weight, attractiveness balance and tie balance by investigating their correlations with ECN size. Without loss of generality, the data of January is analyzed in the following experiments, and the rest of data are used for verifying the results obtained. All the methods in processing and presenting data are the same for $\overline{w}$, $\eta$ and $\theta$. Thus, we only describe these methods in studying the correlation between $\overline{w}$ and $k^{out}$, and discuss the results for all of them.

Initially, we investigate the correlation between average node weight $\overline{w}$ and ECN size. In Fig. \ref{Fig1}(a), the 2-dimensional distribution of $k^{out}$ and $\overline{w}$ is illustrated. As the data points with the same $k^{out}$ are concentrated, we use the mean value of $\overline{w}$ to represent all the data points with the same $k^{out}$. The correlation between $\overline{w}$ and $k^{out}$ is illustrated in Fig. \ref{Fig1}(b). It can be seen that $\overline{w}$ and $k^{out}$ are not related before some $k^{out}$, suggesting that people tend to have the same level of average emotional closeness with their alters regardless of their ECN sizes. Nevertheless, with the expanding of ECN, they are becoming negatively correlated, meaning that the average emotional closeness between ego and the alters decreases with the increase of ECN size. Taken all together, when en ego has a smaller ECN than $k_{c}^{out}$, ego have enough time and resources to maintain the emotional closeness with the alters. Once the ECN size surpasses $k_{c}^{out}$, the ego cannot maintain the the same level of close relationships, resulting in the decrease of the average node strength. In order to identify such critical size, we calculate $\rho^{\overline{w}}$ and show the results in Fig. \ref{Fig1}(c). $\rho^{\overline{w}}$ is high at first for lacking of data points. With the increase of $k^{out}$, $\rho^{\overline{w}}$ approaches the correlation between $\overline{w}$ and $k^{out}$, thus it drops when $k^{out}$ is less than $k_{c}^{out}$. As $k^{out}$ goes beyond $k_{c}^{out}$, the correlation increases. As a result, the minimum value corresponds to $k_{c}^{out}$. For $\overline{w}$, the critical size of ECN is 161.

\begin{figure*}[!tbp]
\centering
\includegraphics[width=0.9\textwidth]{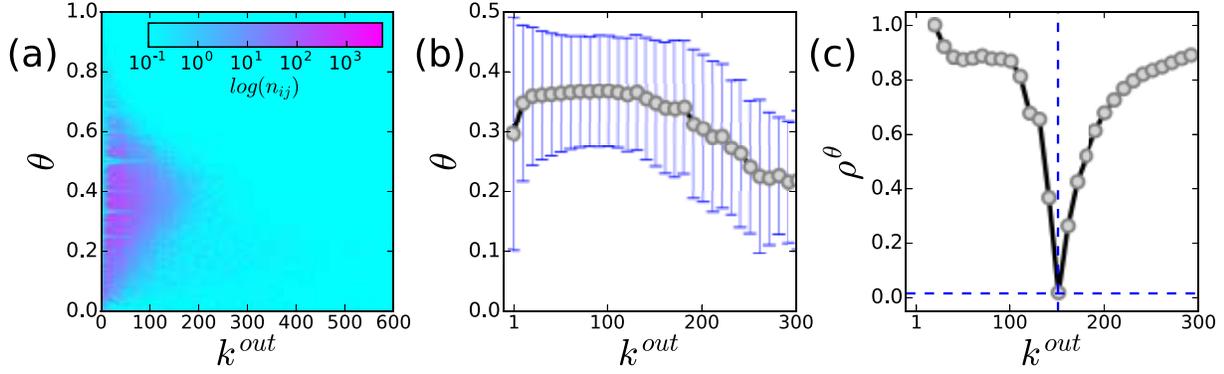}
\caption{(Color online) 2-dimensional distribution and correlation for tie balance $\theta$ and ECN size $k^{out}$. (a) 2-dimensional distribution for $\theta$ and $k^{out}$. The color represents the number of egos ($n_{ij}$) in log scale. (b) Correlation between $\theta$ and $k^{out}$. The data are divided into bins with length 10 and the average value of each bin is presented. The error bar corresponds to the standard deviation. (c) The critical size $k_{c}^{out}$ is identified as 151, where $\rho^{\theta}$ reaches its minimum value.}
\label{Fig3}
\end{figure*}

Furthermore, we investigate the correlation between attractiveness balance and ECN size. In Fig. \ref{Fig2}(a), the 2-dimensional distribution of $k^{out}$ and $\eta$ is illustrated. We investigate the correlation between $\eta$ and $k^{out}$. The results are illustrated in Fig. \ref{Fig2}(b). The same pattern from Fig. \ref{Fig1}(b) can be found here, where the attractiveness balance between ego and the network keeps at 1 before the critical ECN size, meaning most people can maintain the balance relationship with the network. With the ECN size goes on increasing, people tend to have less in-coming contacts. In general, most egos distribute their time and resources in maintaining a balanced relationship when the ECN size is smaller than $k_{c}^{out}$, once their ECNs exceed $k_{c}^{out}$, egos have to decrease their attentions spending on answering the in-coming contacts. In order to find the exact critical ECN size, we calculating $\rho^{\eta}$ and show the results in Fig. \ref{Fig2}(c). Considering the attractiveness balance, the critical size of ECN here is 121.

Last but not least, we study the correlations between tie balance and ECN size. In Fig. \ref{Fig3}(a), the 2-dimensional distribution of $k^{out}$ and $\theta$ is illustrated. The correlation diagram is demonstrated in Fig. \ref{Fig3}(b), and the same patterns from Fig. \ref{Fig1}(b) and Fig. \ref{Fig2}(b) appear again. Within the critical size of ECN, egos can preserve a steady proportion of strong ties regardless of the ECN size. When egos go on increasing their ECN sizes, such proportions will drop significantly. This means egos usually tend to keep a balanced proportion of strong and weak ties if they have enough time and resources in maintaining social relationships, once the resources are used up, they have to decrease the strong ties as the expense of increasing the ECN sizes. In order to identify the exact value of $k_{c}^{out}$, we calculate $\rho^{\theta}$ and the result is presented in Fig. \ref{Fig3}(c). The critical size turns out to be 151 for tie balance.

\begin{figure}[!tbp]
\begin{center}
\includegraphics[width=0.42\textwidth]{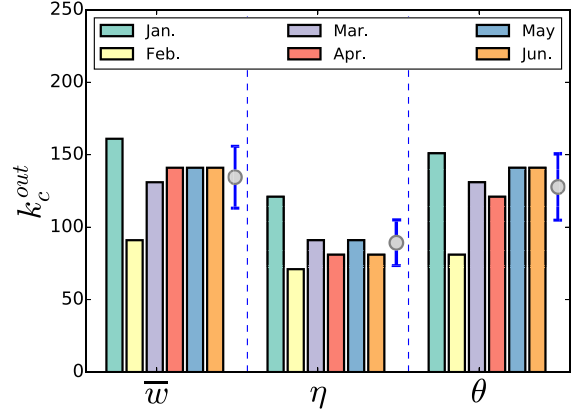}
\caption{(Color online) Critical sizes $k_{c}^{out}$ obtained from the data of six months. The color bars mark the critical sizes of a metric in different months. The dot and the error bar represent the mean and standard deviation of the $k_{c}^{out}$ for each metric, respectively. All the $k_{c}^{out}$s are close to 150, which is the well-known ¡°Dunbar's Number¡±.}
\label{Fig4}
\end{center}
\end{figure}

For a brief summary, the ECN properties are not related to its size when it is less than $k_{c}^{out}$. This means when the ECN size are smaller than $k_{c}^{out}$, people tend to organize their ECNs in a similar way with the same level of average emotional closeness and structural balance (attractiveness balance and tie balance). Once the ECN size goes beyond $k_{c}^{out}$, the average emotional closeness drops and the balanced structure collapses. $k_{c}^{out}$s are obtained by calculating $\rho^{\overline{w}}$, $\rho^{\eta}$ and $\rho^{\theta}$. The results show consistency of about 150, which is close to the well known ``Dunbar's Number". In order to verify our findings, we further calculate $k_{c}^{out}$s for the other five months from February to June, 2014. The results are shown in Fig.~\ref{Fig4}. It can be seen that the critical sizes of ECNs persist over time, and they are close to the well-known ``Dunbar's Number". For a more intuitive understanding, egos have to spend more cognitive resources in maintaining their relationships as the ECN size increases. However, the total amount of cognitive resources for an ego is limited, resulting in the reduction of ego's average emotional closeness with alters as well as the proportion of reciprocal contacts. All these empirical results suggest that ECN size plays a crucial role in its organization and properties.

\section{Conclusion and discussions}

In summary, we have studied how the ECN size affects the properties of the mobile communication networks from ego perspective by investigating the correlations between average node weight, attractiveness balance, tie balance and the ECN size based on the directed graph built from CDR data of about 7 million users. The results showed that once the ECN exceeds a critical size, the emotional closeness drops significantly, the balanced relationship between an ego and the network collapses and the proportion of strong ties decreases. The critical sizes obtained from the data of six months with different ECN metrics showed consistency with ``Dunbar's Number", which claimed that such limitations came from the neocortex size of human themselves. The application of the directed model and the communication data of a city distinguish this research from the related works. Our findings can be viewed as a cross-culture supportive evidence for the SBH theory.

In this paper, we focus on the relationships between the ego and his/her direct contacts. In fact, the relationships among his/her contacts also play a crucial role in understanding the structures of ego networks, thus they need further investigations. As for future work, it is interesting to explore the network structures of ECNs, considering both link formations and node properties. Moreover, the rapid development of information technologies makes it possible to record more dimensions of personal communication data, thus enabling us to achieve deeper understanding in how people organize their ego networks.

\acknowledgments
The authors acknowledge Zhihai Rong and Liming Pan for useful discussions. This work was partially supported by the National Natural Science Foundation of China (Grant Nos. 61302077, 61421061, 61433014, 11222543 and 61502083) and the Key Project of the National Research Program of China (Grant No. 2012BAH41F03). Qing Wang acknowledges the joint cooperation project of BUPT and China Telecom Beijing Institute.


\begin{thebibliography}{10}
\bibitem{Borgatti2009}
\Name{Borgatti S. P., Mehra A., Brass D. J. \and Labianca G.}
\REVIEW{Science}{323}{2009}{892}.
\bibitem{Zhao2011}
\Name{Zhao Z.-D., Xia H., Shang M.-S. \and Zhao T.}
\REVIEW{Chin. Phys. Lett.}{28}{2011}{068901}.
\bibitem{Fisher2005}
\Name{Fisher D.}
\REVIEW{IEEE Internet Comput.}{9}{2005}{20}.
\bibitem{Gonzalez2008}
\Name{Gonzalez M. C., Hidalgo C. A. \and Barab{\'a}si A.-L.}
\REVIEW{Nature}{453}{2008}{779}.
\bibitem{Song2010}
\Name{Song C., Qu Z., Blumm N. \and Barab{\'a}si A.-L.}
\REVIEW{Science}{327}{2010}{1018}.
\bibitem{Miritello2011}
\Name{Miritello G., Moro E., Lara R.}
\REVIEW{Phys. Rev. E}{83}{2011}{045102}.
\bibitem{Toole2015}
\Name{Toole J. L., Herrera-Yaq{\"u}e C., Schneider C. M. \and Gonz{\'a}lez, M. C.}
\REVIEW{J. R. Soc. Interface}{105}{2015}{20141128}.
\bibitem{Roberts2011}
\Name{Roberts S. G. B. \and Dunbar R. I. M.}
\REVIEW{Personal Relationships}{18}{2011}{439}.
\bibitem{Arnaboldi2012}
\Name{Arnaboldi V., Conti M., Passarella A. \and Pezzoni F.} in
\Book{IEEE International Conference on Social Computing (SocialCom)} (IEEE Press)
\Year{2012}\Pages{31}{40}.
\bibitem{Onnela2007}
\Name{Onnela J.-P., Saram{\"a}ki J., Hyv{\"o}nen J., Szab{\'o} G., Lazer D., Kaski K., Kert{\'e}sz J. \and Barab{\'a}si A.-L.}
\REVIEW{Proc. Natl. Acad. Sci. USA}{104}{2007}{7332}.
\bibitem{Nanavati2008}
\Name{Nanavati A. A., Singh R., Chakraborty D., Dasgupta K., Mukherjea S., Das G., Gurumurthy S. \and Joshi A.}
\REVIEW{IEEE Trans. Knowl. Data Eng.}{20}{2008}{703}.
\bibitem{Eagle2009}
\Name{Eagle N., Pentland A. S. \and Lazer D.}
\REVIEW{Proc. Natl. Acad. Sci. USA}{106}{2009}{15274}.
\bibitem{Wu2010}
\Name{Wu Y., Zhou C., Xiao J., Kurths J. \and Schellnhuber H. J.}
\REVIEW{Proc. Natl. Acad. Sci. USA}{107}{2010}{18803}.
\bibitem{Jiang2013}
\Name{Jiang Z.-Q., Xie W.-J., Li M.-X., Podobnik B., Zhou W.-X. \and Stanley H. E.}
\REVIEW{Proc. Natl. Acad. Sci. USA}{110}{2013}{1600}.
\bibitem{Miritello2013}
\Name{Miritello G., Moro E., Lara R., Mart{\'\i}nez-L{\'o}pez R., Belchamber J., Roberts S. G. B. \and Dunbar R. I. M}
\REVIEW{Soc. Networks}{35}{2013}{89}.
\bibitem{Miritello2013b}
\Name{Miritello G., Lara R., Cebrian M. \and Moro E.}
\REVIEW{Sci. Rep.}{3}{2013}{1950}.
\bibitem{Saramaki2014}
\Name{Saram{\"a}ki J., Leicht E. A., L{\'o}pez E., Roberts S. G. B., Reed-Tsochas F. \and Dunbar R. I. M.}
\REVIEW{Proc. Natl. Acad. Sci. USA}{111}{2014}{942}.
\bibitem{Dunbar1998}
\Name{Dunbar R. I. M.}
\REVIEW{Evolutionary Anthropology}{6}{1998}{178}.
\bibitem{Dunbar1992}
\Name{Dunbar R. I. M.}
\REVIEW{Journal of Human Evolution}{22}{1992}{469}.
\bibitem{Dunbar2010}
\Name{Dunbar R. I. M.}
\Book{How Many Friends Does One Person Need? Dunbar¡¯s Number and Other Evolutionary Quirks}
\Editor{Julian Loose} \Publ{Faber \& Faber, Cambridge}
\Year{2010} \Pages{21}{43}.
\bibitem{Ruiter2011}
\Name{Ruiter J. D., Weston G. \and Lyon S. M.}
\REVIEW{American Anthropologist}{113}{2011}{557}.
\bibitem{Arnaboldi2013}
\Name{Arnaboldi V., Conti M., Passarella A. \and Pezzoni F.} in
\Book{The 5th IEEE International Workshop on Network Science for Communication Networks (NetSciCom)} (IEEE Press)
\Year{2013}\Pages{229}{234}.
\bibitem{Dunbar2015}
\Name{Dunbar R. I. M., Arnaboldi V., Conti M. \and Passarella A.}
\REVIEW{Soc. Networks}{43}{2015}{39}.
\bibitem{Zhao2014}
\Name{Zhao J.-C., Wu J.-J., Liu G.-N., Tao D.-C., Xu K. \and Liu C.-Y.}
\REVIEW{Neurocomputing}{142}{2014}{343}.
\bibitem{Gonccalves2011}
\Name{Gon{\c{c}}alves B., Perra N. \and Vespignani A.}
\REVIEW{PLoS ONE}{6}{2011}{e22656}.
\bibitem{Guo2013}
\Name{Guo Q., Shao F., Hu Z.-L. \and Liu J.-G.}
\REVIEW{EPL}{104}{2013}{28004}.
\bibitem{Zhou2005}
\Name{Zhou W.-X., Sornette D., Hill R. A. \and Dunbar R. I. M.}
\REVIEW{Proc. R. Soc. London Ser. B}{272}{2005}{439}.
\bibitem{Brzozowski2011}
\Name{Brzozowski M. J. \and Romero D. M.} in
\Book{Fifth International AAAI Conference on Weblogs and Social Media (ICWSM)} (AAAI Press)
\Year{2011}\Pages{458}{461}.
\bibitem{Brown1987}
\Name{Brown J. J. \and Reingen P. H.}
\REVIEW{J. Consum. Res.}{14}{1987}{350}.
\bibitem{Holt2010}
\Name{Holt-Lunstad J., Smith T. B. \and Layton J. B.}
\REVIEW{PLoS Medicine}{7}{2010}{e1000316}.
\bibitem{Levandowsky1971}
\Name{Levandowsky M.\and Winter D.}
\REVIEW{Nature}{234}{1971}{34}.
\bibitem{Pearson1895}
\Name{Pearson K.}
\REVIEW{Proc. Roy. Soc. Lond.}{58}{1895}{240}.
\end{thebibliography}
\end{document}